\documentclass{PoS}

\title{Parton showers with medium-modified splitting functions}

\ShortTitle{Parton showers with medium-modified splitting functions}

\author{\speaker{Gennaro Corcella}\\
  Museo Storico della Fisica, Centro Studi e Ricerche E.~Fermi,\\
             Piazza del Viminale 1, I-00184 Roma, Italy,\\   
Scuola Normale Superiore, Piazza dei Cavalieri 7, I-56123 Pisa, Italy,\\
INFN, Sezione di Pisa,  Largo Fibonacci 3,
I-56127, Pisa, Italy\\      
        E-mail: \email{gennaro.corcella@sns.it}}

\abstract{I discuss the recent implementation of medium-modified
splitting functions within the HERWIG angular-ordered parton shower
algorithm and present a few results on transverse momentum, 
energy and angular distributions.}

\FullConference{XVIII International Workshop on Deep-Inelastic Scattering and Related Subjects, DIS 2010\\
		April 19-23, 2010\\
		Firenze, Italy}

\begin{document}
Much work has been carried out through the years in order to theoretically
describe jet quenching, one of the most striking observations of heavy-ion
collisions, namely the suppression of 
particle multiplicity at large transverse momentum
($p_T$),
as well as other related phenomena,
such as the disappearance or distortion of the spectra of particles at opposite
directions with respect to a reference one \cite{rhic}.
A higher energy loss in a dense medium is a possible explanation of jet quenching:
partons which in the vacuum are potentially produced at large $p_T$ can
further radiate, which determines suppression of the high-$p_T$ spectrum
and enhancement of the low-$p_T$ one.

In order to describe the higher branching probability in a medium,
a simple prescription consists in adding to the Altarelli--Parisi
splitting functions a term depending on the medium 
properties \cite{pol,mmff},
i.e. 
\begin{equation}\label{dp}
P(z)\to P(z)+\Delta P(z,p^2,E,\hat q,L).
\end{equation}
In (\ref{dp}), $p^2$ is the branching-parton virtuality, $E$ its energy,
$L$ the medium length, $\hat q$ the transport coefficient, defined as
the average transverse momentum transferred from the medium to the
parton per unity of free path.
The correction $\Delta P(z,p^2,E,\hat q,L)$ can be computed by means of
the so-called BDMPS approximation \cite{dok}, namely multiple collinear
radiation off static scattering centres, in the presence of
a screened Coulomb potential.

Such modified splitting functions can be implemented in
Monte Carlo parton shower generators, which can thus be used by the heavy-ion
community to compare with data from RHIC and, ultimately, LHC experiments.
Medium modifications have been included in the framework of the
PYTHIA event generator \cite{pythia}, which
orders cascades in virtuality, with an option to
to veto branchings which do not satisfy angular ordering.
This implementation is known as Q-PYTHIA \cite{acs,qp}.
More recently \cite{accs}, medium-modified splittings have also been implemented 
in the HERWIG showering algorithm \cite{hw}, which, unlike PYTHIA, systematically
includes angular ordering, thus leading to colour coherence in the
large-$N_C$ limit.
Following \cite{accs}, herafter I shall present the main results yielded by
medium-modified HERWIG parton showers.

For the sake of comparison with the Q-PYTHIA results in \cite{acs}, 
I shall study the cascades initiated by a single gluon, whose
energy will be fixed to $E=10$ and 100 GeV, with the hadronization
switched off.
As discussed in \cite{accs},
since this is not a standard HERWIG option, it was necessary to add a
fictitious process and set by the hand the upper limit of the
evolution variable $Q^2\simeq E^2(1-\cos\theta)$ to 
$Q^2_{\mathrm{max}}=2E^2$, where 
$E$ is the energy of the branching parton and $\theta$ the emission
angle. 
Morover, care must be taken about the fact that the medium
length varies along the shower: in fact, if 
$L_0$ is the length for the first splitting, 
in a subsequent emission, a parton `sees' an effective length
$L=L_0-2 z E/k_T^2$, $z$ and $k_T$ being its energy fraction and
transverse momentum, since it  
has travelled for a distance $2 z E/k_T^2$, the so-called parton formation
length, before radiating again.
As $L$ is not positive definite, 
whenever the medium length becomes negative, the shower will 
be vacuum-like.

Hereafter, I shall consider the options of media with
$\hat q=$~1 and 10~GeV$^2$/fm, $L_0=2$ and 5~fm. For simplicity,
such medium configurations will be labelled 
in terms of the so-called 
accumulated transverse momentum $\hat qL_0=2$,
5, 20 and 50 GeV$^2$.

\begin{figure}
\centerline{\resizebox{0.49\textwidth}{!}{\includegraphics{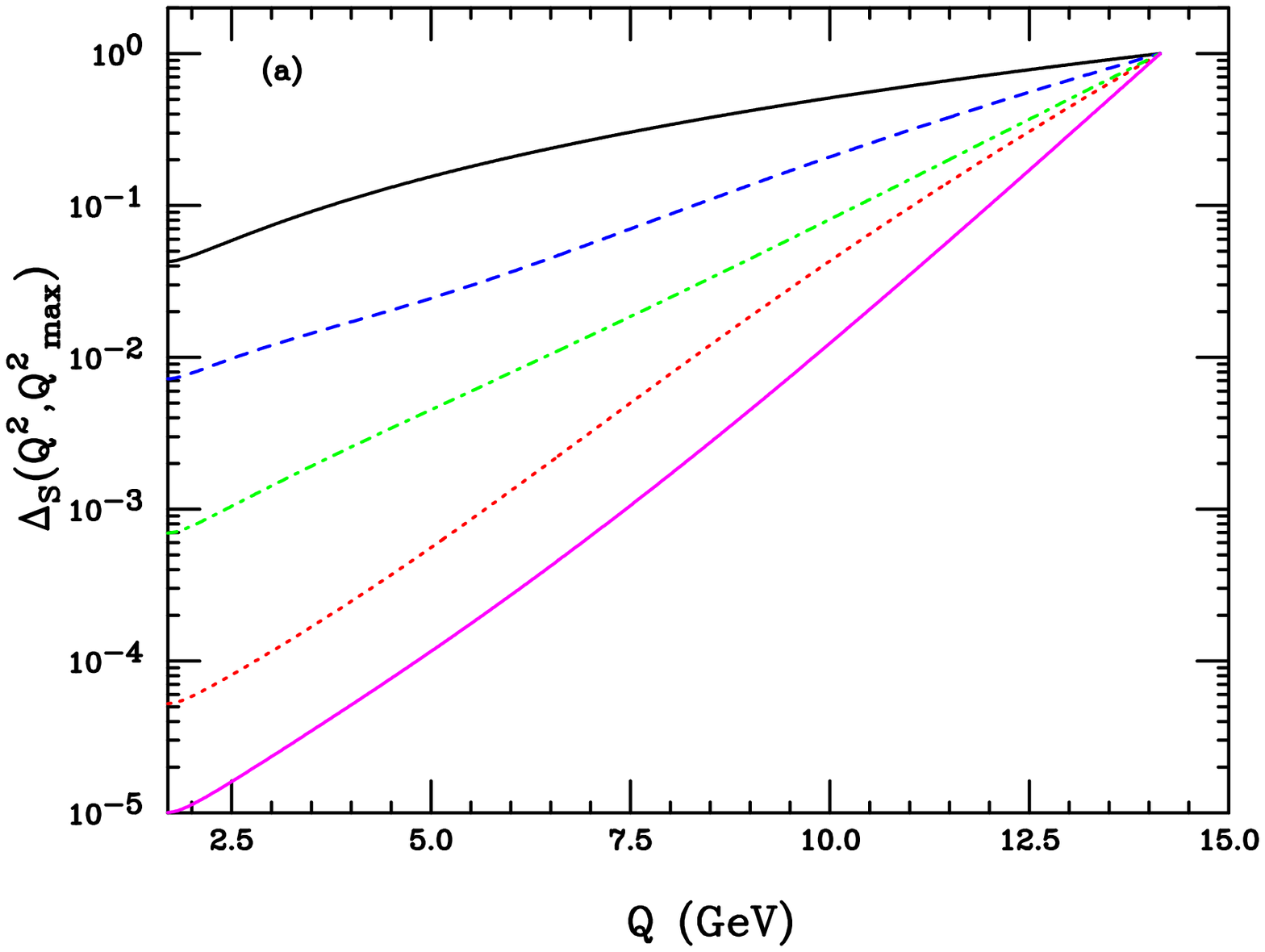}}%
\hfill%
\resizebox{0.49\textwidth}{!}{\includegraphics{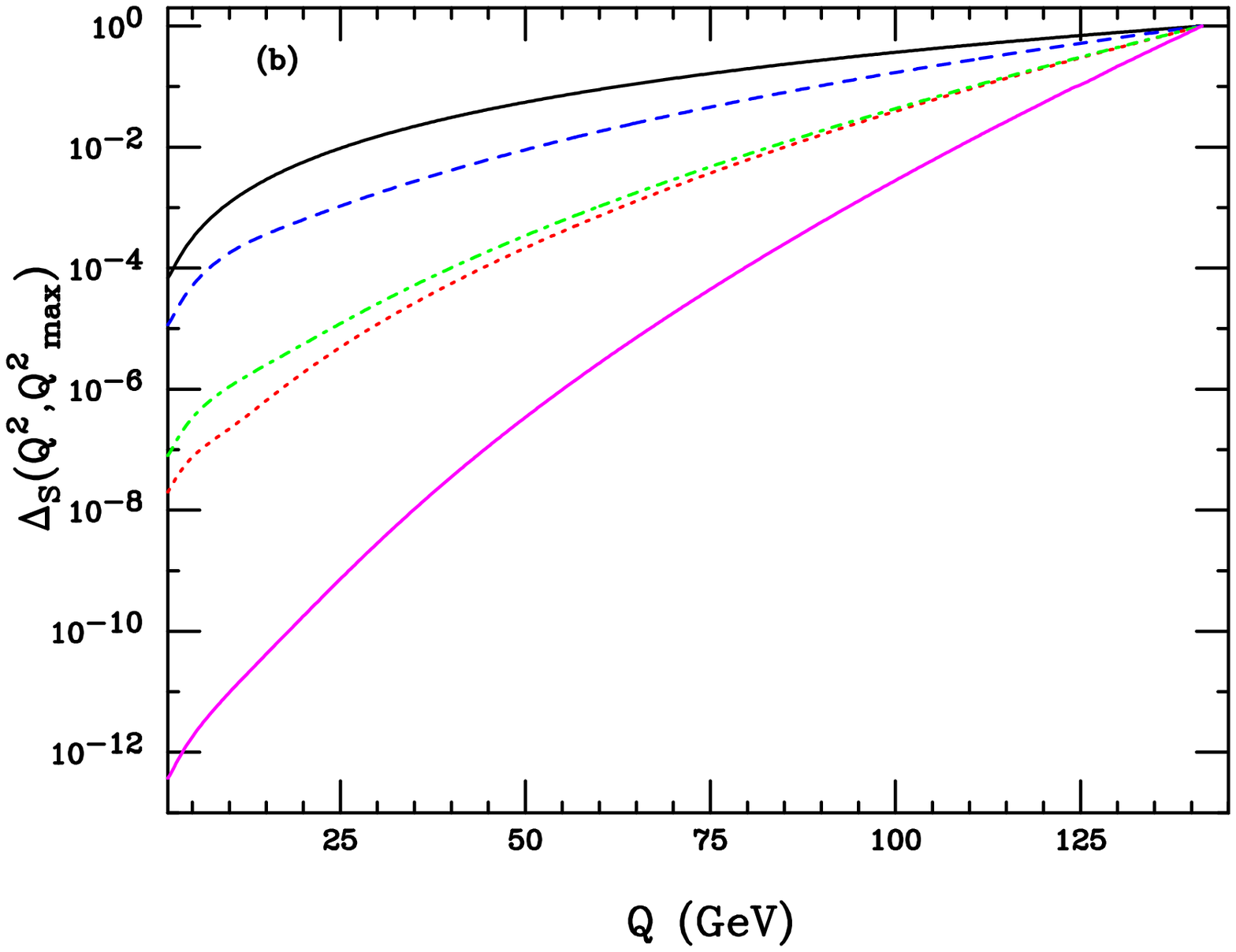}}}
\caption{Gluon Sudakov form factors, in the vacuum (solid, black) and
in media with $\hat qL_0=2$ (dashes, blue), 5
(dots, red), 20 (dot-dashes, green) and 50 (magenta, solid) GeV$^2$.
The gluon energy is $E=10$~GeV (a)
and 100 GeV (b).}
\label{sud}
\end{figure}

A fundamental quantity in shower algorithms is the Sudakov form
factor $\Delta_S(Q_1^2,Q_2^2)$, namely the probability of evolution
between $Q_1^2$ and $Q_2^2$ with no intermediate branching.
Fig.~\ref{sud} presents the gluon Sudakov form factors
$\Delta_S(Q^2,Q_\mathrm{max}^2)$, 
for $E=10$ and 100 GeV, fixed $L=L_0$ and the values of
$\hat q$ and $L_0$ given above. 
From Fig.~\ref{sud} one learns that  
medium-induced effects are quite large: the Sudakov form factors exhibit
suppressions of a few orders of magnitude, with respect to the vacuum,
corresponding to an enhancement of the branching probability, and thus
of the radiative energy loss.

\begin{figure}[t]
\centerline{\resizebox{0.49\textwidth}{!}{\includegraphics{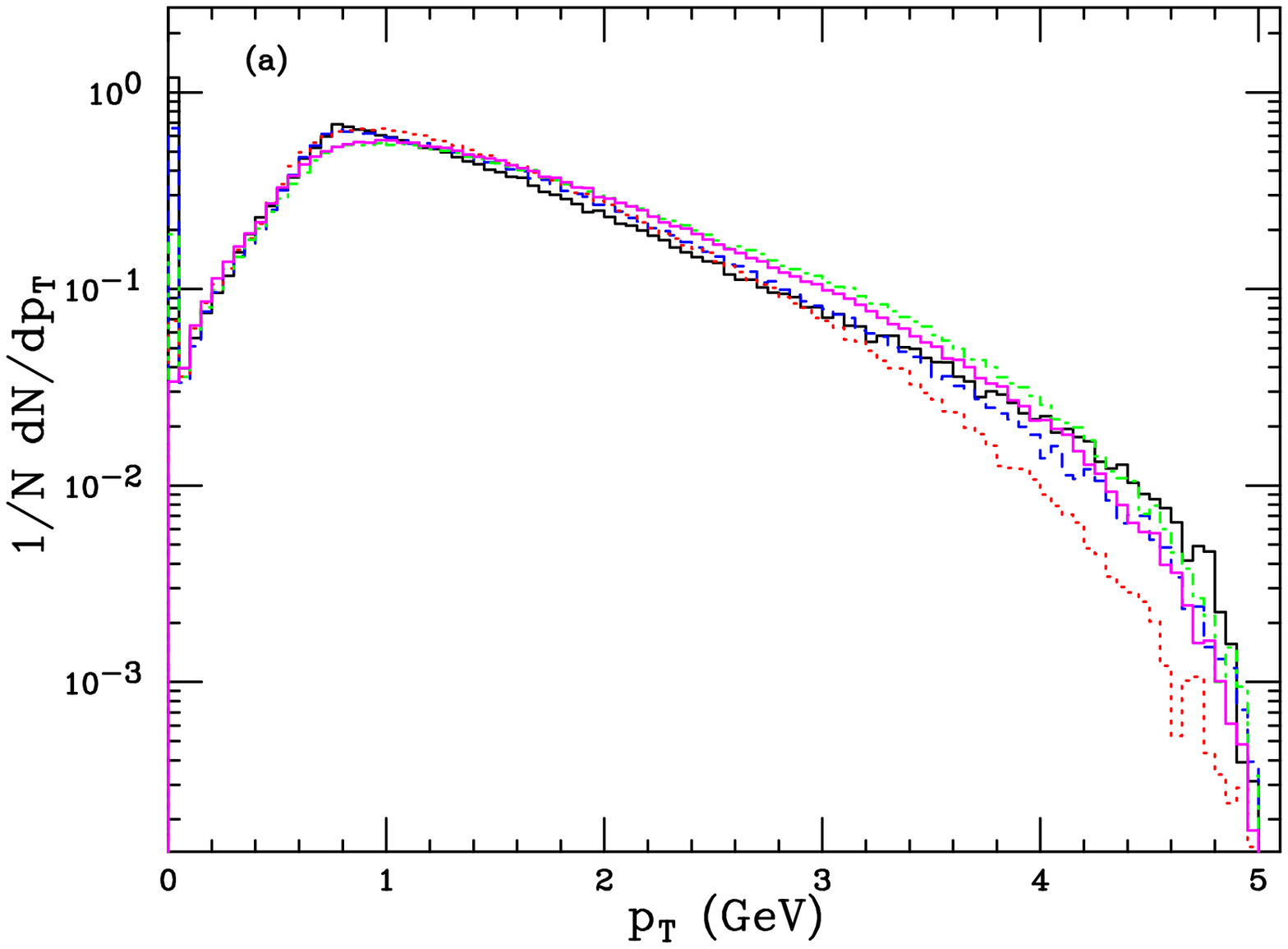}}%
\hfill%
\resizebox{0.49\textwidth}{!}{\includegraphics{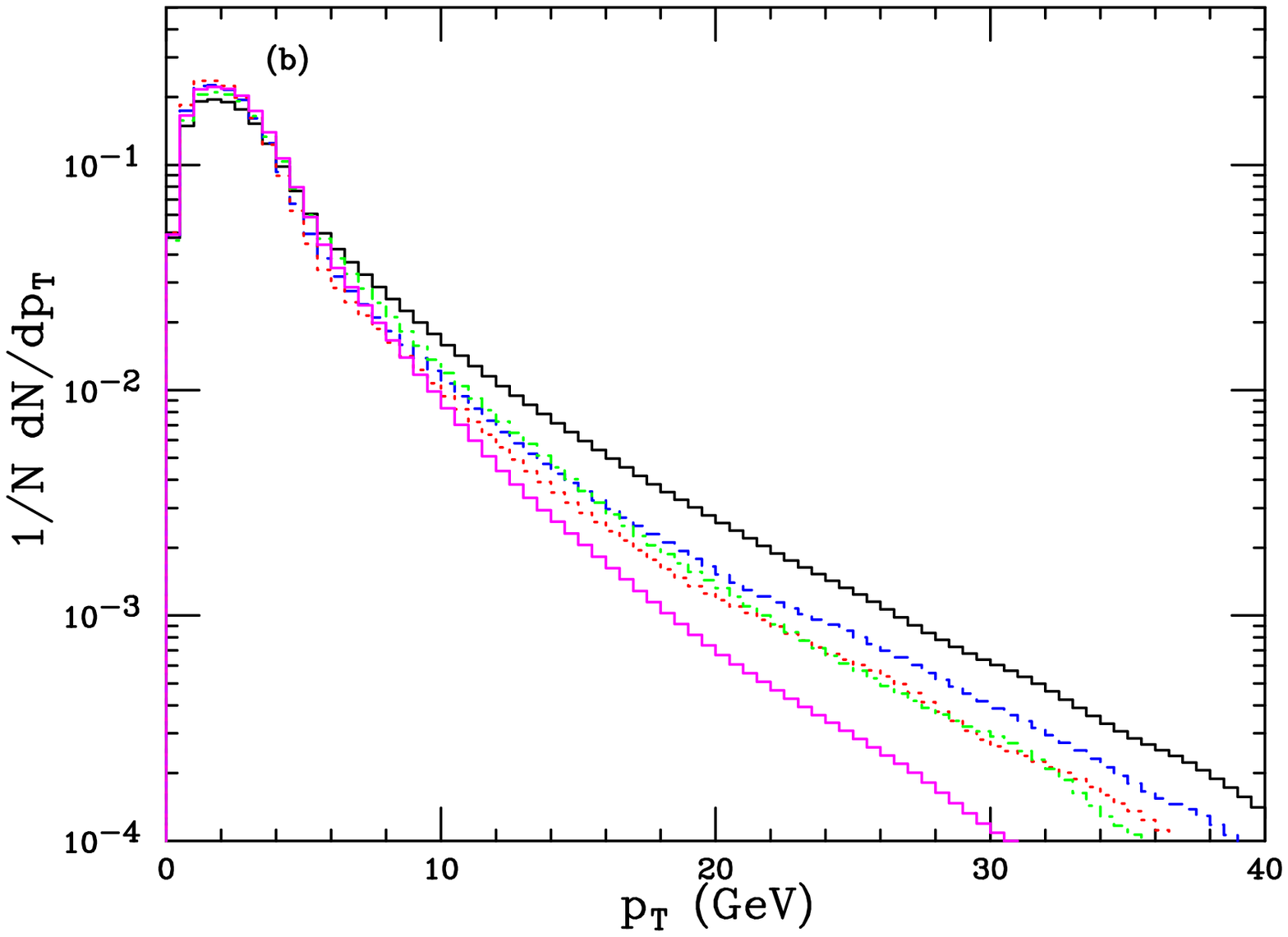}}}
\caption{Transverse momentum multiplicity in a  
shower initiated by a gluon of energy 10 GeV (a) and 100 GeV (b),
in the vacuum (solid, black) and in media with accumulated
transverse momentum $\hat qL_0=2$ (dashes, blue), 
5 (dots, red), 20 (dot-dashes, green) and 50 (solid, magenta) GeV$^2$.}
\label{pt}
\end{figure}

\begin{figure}
\centerline{\resizebox{0.49\textwidth}{!}{\includegraphics{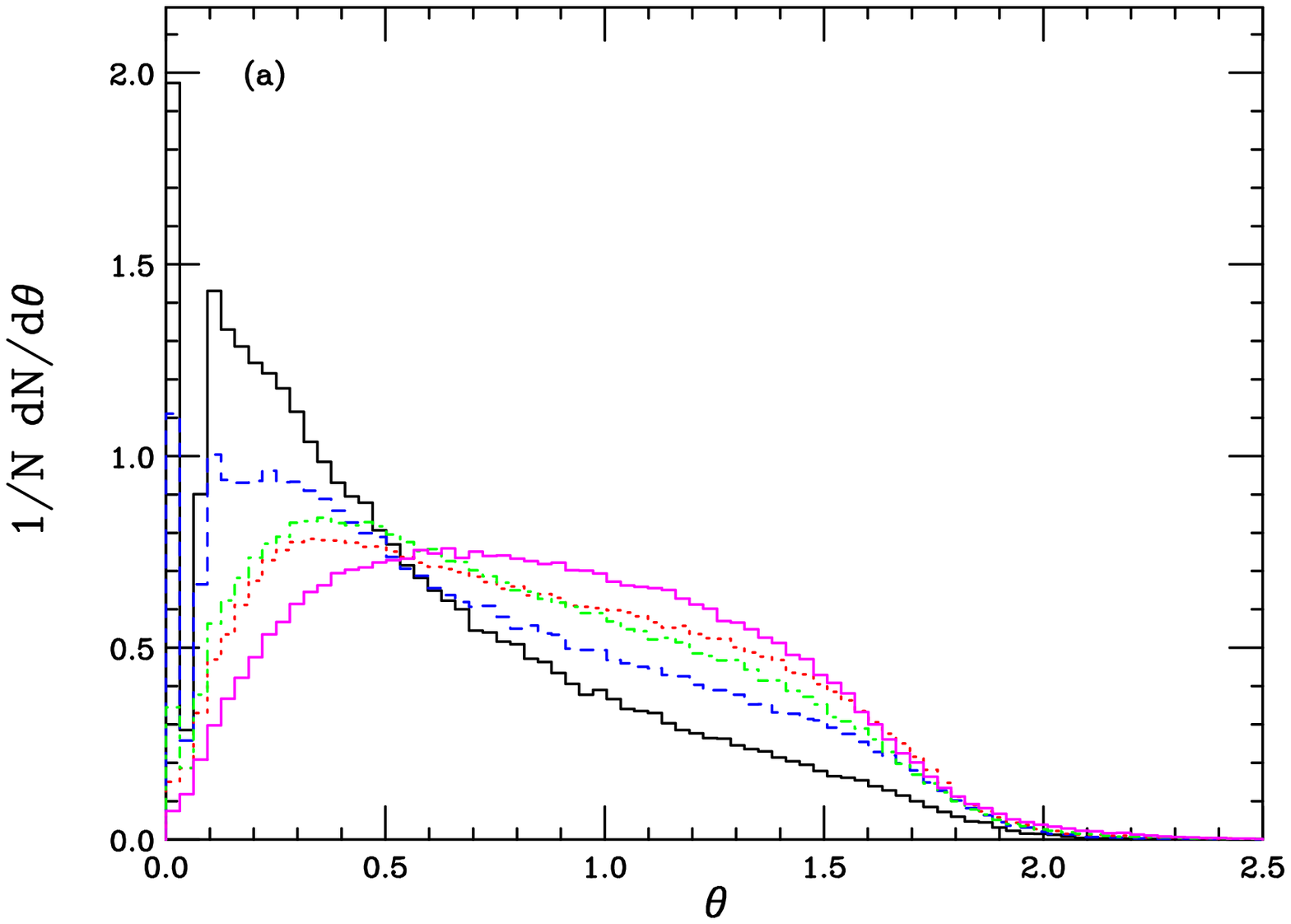}}%
\hfill%
\resizebox{0.49\textwidth}{!}{\includegraphics{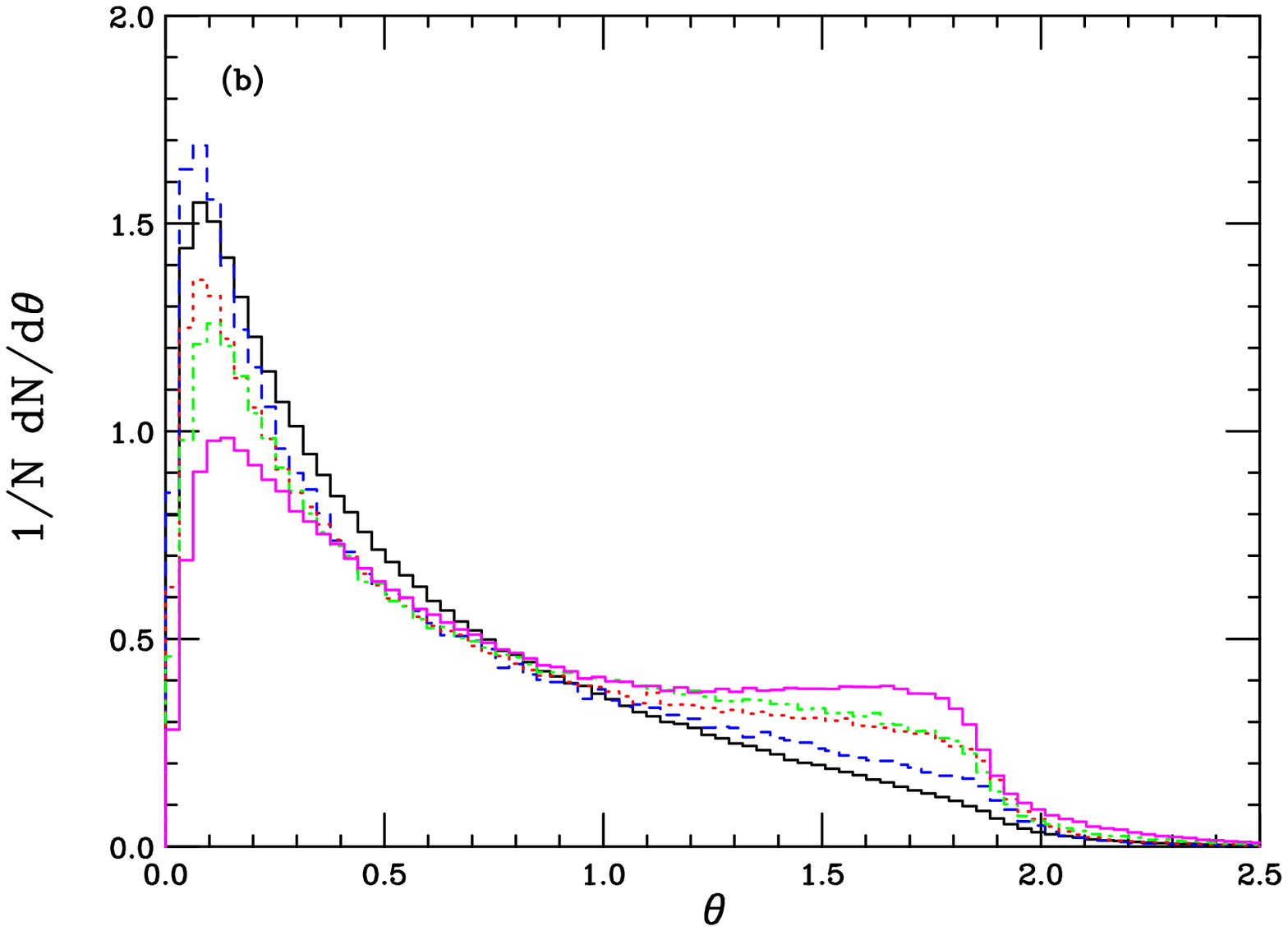}}}
\caption{
Angular distributions for showers 
initiated by a gluon of energy 10 GeV (a) and 100 GeV (b).
The lines are labelled as in Fig.~2.}
\label{teta}
\end{figure}

As a further check of the higher emission probability, 
Table~\ref{mult} quotes the 
the average parton multiplicity in the vacuum
and in a dense medium. The enhancement factor
runs from  20\% ($\hat qL_0=2$~GeV$^2$) 
up to about 80\% ($\hat qL_0=50$~GeV$^2$) for $E=10$~GeV, and
between 7\% and 70\% for 100 GeV.
\begin{table}
\small
\caption{\label{mult}
Average parton multiplicities in showers initiated by gluons 
of energy of 10 and
100 GeV, in the vacuum and in a medium with assigned values of $\hat qL_0$.}
\begin{center}
\begin{tabular}{| c || c | c | c | c | c ||}
\hline
$E$ & $\hat q L_0=0$ & 
$\hat qL_0=2$~GeV$^2$ & $\hat qL_0=5$~GeV$^2$ & 
$\hat qL_0=20$~GeV$^2$ & $\hat qL_0=50$~GeV$^2$ \\
\hline 
10~GeV & 2.56 & 3.05 & 4.14 & 3.60 & 4.56\\
\hline
100~GeV & 6.95 & 7.41 & 8.79 & 8.93 & 11.70\\
\hline
\end{tabular}
\end{center}
\end{table}  
In Figs.~\ref{pt}--\ref{teta} I present the differential multiplicity
with respect to the parton transverse momentum ($p_T$),
angle ($\theta$) and logathmic energy fraction, defined as 
$\xi=\ln (E/|p|)$, where $E$ is the initiating-gluon energy
and $|p|$ the modulus of the three-momentum of the partons in the cascade.
All plotted distributions are normalized to unity.

Overall, the medium-modified
spectra present the features which one should expect and
in agreement with the jet-quenching observations: suppression
(enhancement) of the parton multiplicity at large (small) $p_T$,
broader angular distributions, small-$\xi$ suppression.
However, compared with the distributions presented in \cite{acs} and obtained
with the Q-PYTHIA code, some differences can be noticed.
As discussed in \cite{accs}, on average the emission probability
in PYTHIA is larger than in HERWIG, as a consequence of 
the different evolution variables and the range wherein they
are allowed to vary.
For example, the vacuum spectra here obtained for $E=10$~GeV exhibit
a sharp peak at $p_T=\theta=\xi=0$, corresponding to a 
fraction of
events wherein the gluon evolves down to the infrared cutoff with no
emission. This feature is less evident when running Q-PYTHIA \cite{acs}.
As for the angular distributions, Fig.~\ref{teta}
shows that the multiplicity
is negligible for $\theta>2$; in fact, unlike PYTHIA,
HERWIG presents a dead zone for large-angle radiation and
we are studying the shower produced by a single parton,
with no matrix-element matching.
Ref.~\cite{accs} also discussed medium-modified showers 
for fixed length $L=L_0$ and reasonable results were obtained,
namely stronger medium effects once $L$ does not decrease 
throughout the cascade.

\begin{figure}[ht!]
\centerline{\resizebox{0.49\textwidth}{!}{\includegraphics{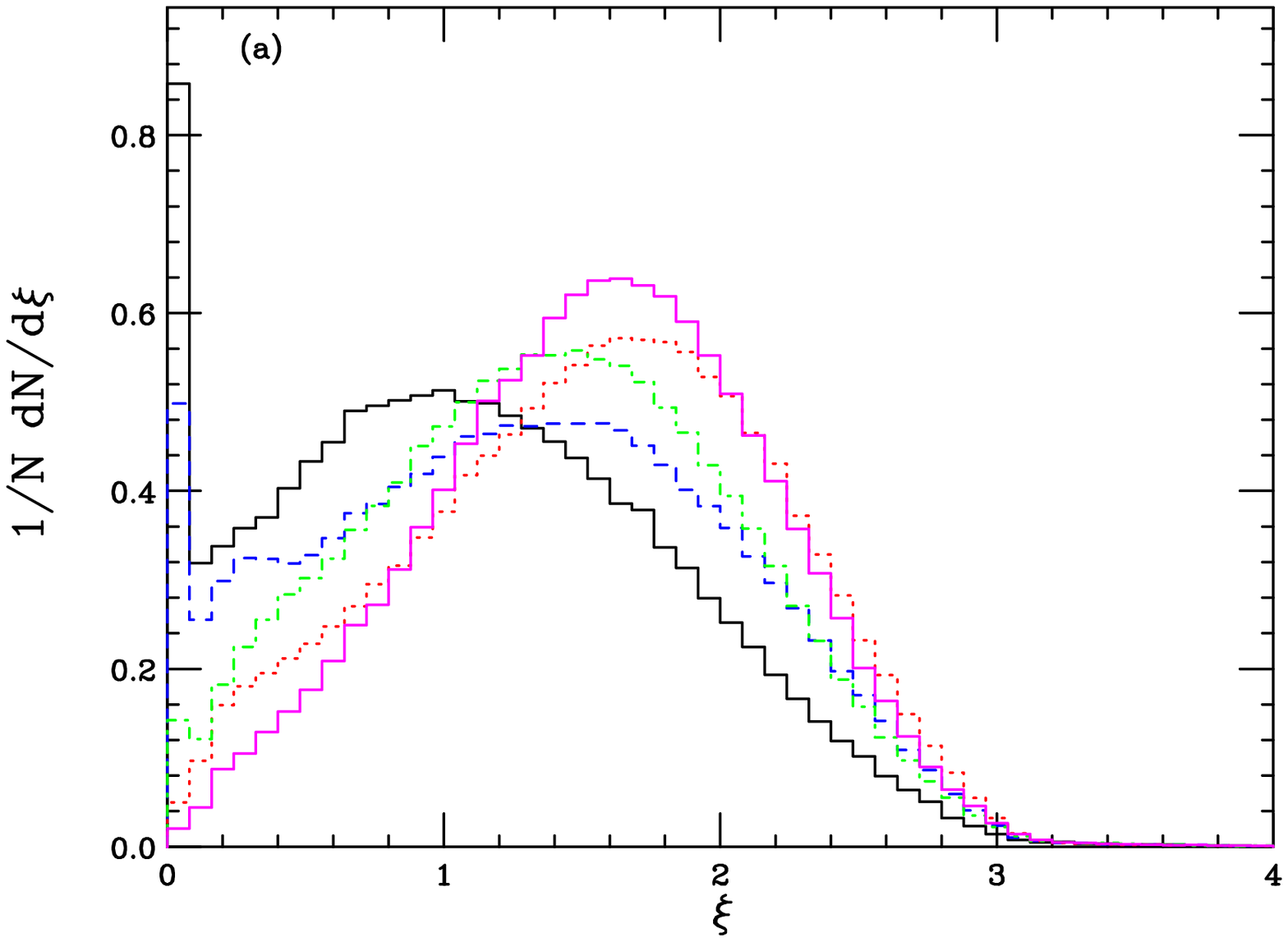}}%
\hfill%
\resizebox{0.49\textwidth}{!}{\includegraphics{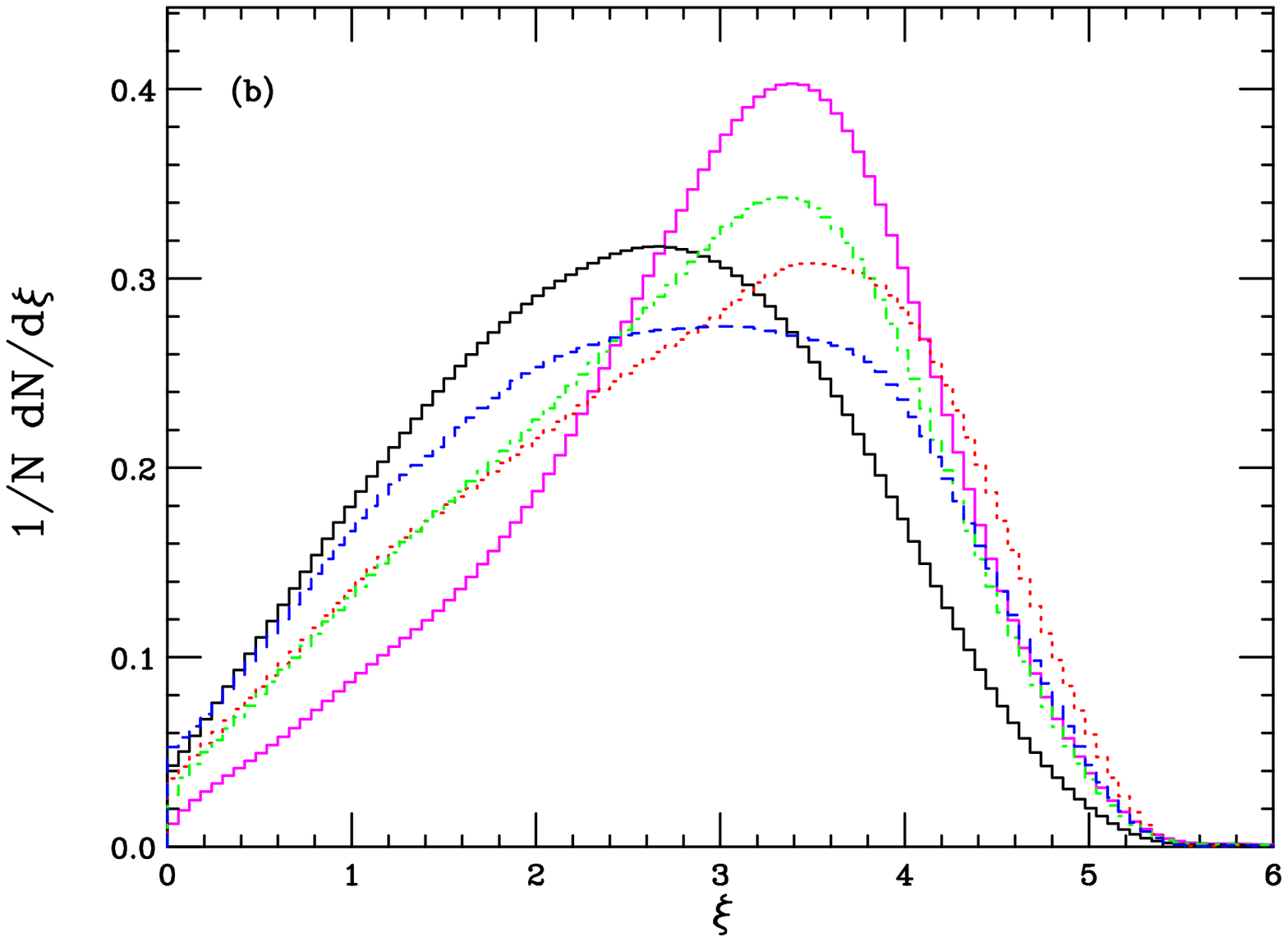}}}
\caption{As in Figs.~2 and 3, but showing
the logarithmic energy-fraction $\xi$.}
\end{figure}

In summary, I reviewed the main issues involved in the modification of
the HERWIG angular-ordered shower algorithm for the sake of including
the effects of a dense medium and possibly describing jet quenching.
The release of such a code, which will be called Q-HERWIG \cite{qh}, is
currently in progress.
Furthermore, in order to draw a final conclusion on the
comparison with Q-PYTHIA,
it will be compelling turning hadronization on and tune
the two programs to the same data set.
In this way, light will be shed even on the role played by 
the evolution variable in medium-modified cascades.

\acknowledgments
{Work carried out in collaboration with N.~Armesto, L.~Cunqueiro and
C.A.~Salgado.}

\end{document}